# Interactive Cycle Model: The Linkage Combination among Automatic Speech Recognition, Large Language Models, and Smart Glasses

**Libo Wang**
Nicolaus Copernicus University
Jurija Gagarina 11, 87-100 Toruń, Poland
326360@o365.stud.umk.pl
UCSI University
Taman Connaught, 56000 Kuala Lumpur, Wilayah Persekutuan Kuala Lumpur, Malaysia
1002265630@ucsi.university.edu.my

## Abstract

This research proposes the interaction loop model "ASR-LLMs-Smart Glasses", which model combines automatic speech recognition, large language model and smart glasses to facilitate seamless human-computer interaction. And the methodology of this research involves decomposing the interaction process into different stages and elements. Speech is captured and processed by ASR, then analyzed and interpreted by LLMs. The results are then transmitted to smart glasses for display. The feedback loop is complete when the user interacts with the displayed data. Mathematical formulas are used to quantify the performance of the model that revolves around core evaluation points: accuracy, coherence, and latency during ASR speech-to-text conversion. The research results are provided theoretically to test and evaluate the feasibility and performance of the model. Detailed architectural details and experimental process have been uploaded to Github, the link is:https://github.com/brucewang123456789/GeniusTrail.git.

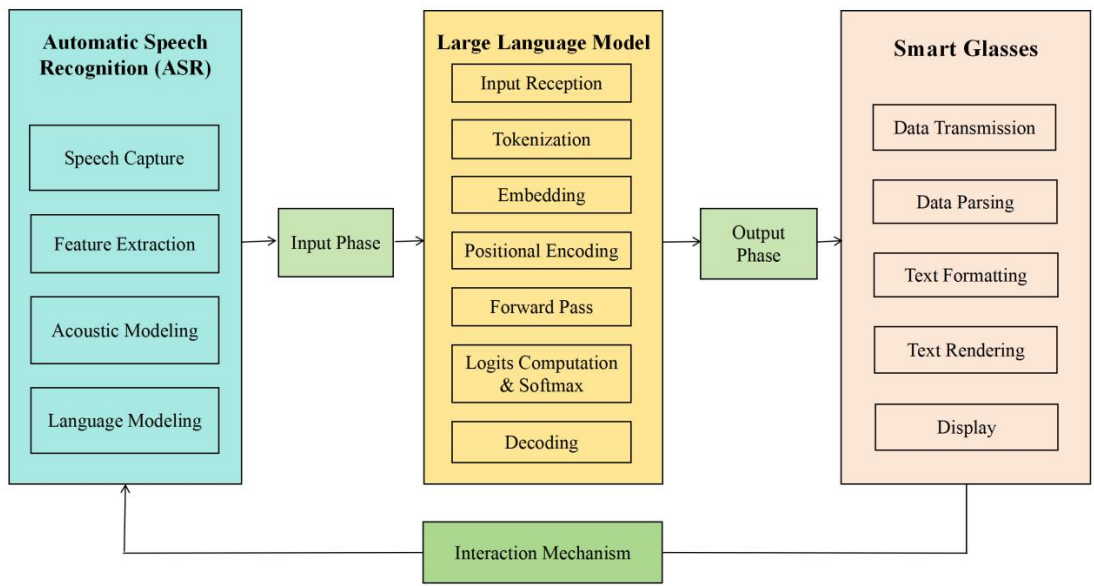

Figure 1 - ASR-LLMs-Smart Glasses Linkage Mechanism

## 1. Introduction

As the complex and rigorous technology in Natural Language Processing (NLP), the essence of Automatic Speech Recognition (ASR) is regarded as relying on technologies such as machine learning and artificial neural networks in many fields, which convert human spoken language into visual written Basic functions of text to facilitate the development of human-computer interaction (HCI) (Li & Hilliges, 2021;

Choutri et al., 2022; Lv et al., 2022). In the initial stage ASR has been able to respond to a limited number of human speech recognition system, and now it has developed into a clear, fluent and accurate complex response system to natural language (Padmanabhan & Johnson Premkumar, 2015).

Starting from speech capture, the operation process of ASR is closely connected with human spoken instructions as the acoustic speech signal it needs to acquire (Cooke et al., 2006). As the GPT series expands to lattice inputs and continues to change, it stimulates the fields of NLP, large language models (LLMs) and Chatbot to gradually recognize accurate and high-quality speech signals and their clear input as the key to the future human-computer interaction systems (Huang & Chen, 2019; Jeon et al., 2023). It follows an adaptive fine-tuning approach for task-specific adaptations and performs adjustments on small task-specific datasets (Hu et al, 2023; Parthasarathy et al., 2024). And for the task of enhancing NLP, LLMs has a more complex neural network and a larger training data set (Dong, 2024).

The integration of ASR technology and LLMs has aroused great interest in industry and academic research after OpenAI developed the advanced ChatGPT, which may realize the current HCI paradigm shift (Tabone & De Winter, 2023). Smart glasses are currently known portable wearable electronic computer devices that can provide enhanced visual effects and text information, which include functions such as integrated displays, cameras, microphones, and wireless connections to smartphones or other host devices (Lee & Hui, 2018). The wearer can not only see the real environment from optical display, but also see the virtual content displayed in the display, which is augmented reality (AR) concept (Kim & Choi, 2021). In recent years, companies such as OpenAI, Meta, and Google have all tried to explore designs and promote smart glasses to the global market, which provides a basis for the practicality of this research (Hou & Bergmann, 2023; Lee et al., 2023).

## 2. ASR-LLMs-Small Glass

Interactive cycle model is a combination of human instinct and artificial intelligence (Figure 1). It uses the linkage of ASR, large language model and smart glasses to strengthen the close relationship between human users and devices. It uses LLMs to generate text and uses ASR to identify human language information to generate what users need. Answers and solutions, and finally visualize them through smart glasses that can both see the real world and possibly see the generated text.

**2.1 Automatic Speech Recognition**

The speech capture stage has a crucial role in ASR that determines how accurately the speech signal can be interpreted and processed by subsequent stages, as a poor quality initial capture can lead to misunderstandings and errors in the final transcription. In the process of converting sound waves into electrical signals.

Microphone arrays are speech capture devices used to convert physical speech signals of air pressure changes into electrical signals (Seltzer, 2003). While the hands-free functionality enables the microphone to intelligently capture not only the meaning of speech desired by the user and LLMs, but also other unwanted noise-causing signals present in the user's location and environment (Wagner et al., 2024).

As shown in previous literature, the final result obtained in the speech capture stage is represented by the digital time series of the speech signal (Nasereddin & Omari, 2017). It needs to preserve the characteristics of the original speech as much as possible, and then extract the characteristics of these digital signals in subsequent stages (Shrawankar & Thakare, 2013).

Feature extraction can apply a window function to the speech signal, which is usually represented as a Hamming window or Hanning window, which divides the continuous speech signal into smaller overlapping frames, which are usually about 20-30 ms long (Ghodasara et al., 2016). When the speech signal is framed, the architecture performs a fast Fourier transform (FFT) for each frame, which converts the signal from the time domain to the frequency domain. Mel spectral coefficients (MFCC) or perceptual linear prediction (PLP) coefficients are usually extracted from frequency domain frames (Langkvist et al., 2014).

Acoustic modeling is a process step in ASR techniques to represent statistical models of the relationship between acoustic signals and speech units (Tachbelie et al., 2014). In modern ASR systems, the most commonly used feature type is the Mel-frequency cepstral coefficient (MFCC) that captures the power spectrum of an audio signal in a manner that approximates the response of the human auditory system (Jin et al., 2013).

Essentially the acoustic modeling task belongs to the pattern recognition problem (Garg & Sharma, 2016) that aims to find the sequence of speech units most likely to produce the observed acoustic features (Chang et al., 2022). This task is usually accomplished by the statistical model of Hidden Markov Model (HMM), because this model is able to process data sequences and incorporate time dependencies between different parts of the speech signal (Monir et al., 2021).

Language modeling is the last key stage of the ASR part of this research. It revolves around the construction of a statistical model that predicts the likelihood of a sequence of words to overcome accuracy challenges due to linguistic uncertainty (Wei et al., 2023).

**2.2 Large Language Model**

LLMs are highly advanced NLP models based on the transformer architecture. Through continuous fine-tuning of the design, it can employ a self-attention mechanism to capture context and construct meaningful text output based on correlation predictions (Chang et al., 2024). Another key element is its positional embedding layer, which assigns tokens to high-dimensional vectors according to their corresponding positions in the input sequence (Hadi et al., 2024).

In the operation process of LLMs, input reception is a crucial link, responsible for receiving text data for further processing, and is directly connected to the language modeling stage of the ASR system (Patel et al., 2022). Tokenization in LLMs is mainly achieved through a technology called byte pair encoding (BPE) (Tavabi & Lerman, 2021). At the heart of the technology lies contextual tokenization, which is processed by splitting textual data into individual units or tokens. This process is driven by BPE, which is designed to cope with infinite vocabularies in a memory-efficient manner (Petrov et al., 2023). Specifically, initial tokens are typically individual characters drawn from a base vocabulary, and longer tokens are then generated by iteratively merging the most frequently adjacent pairs of symbols in the dataset (Xu & Zhou, 2022).

The core functionality of tokenization is that it breaks the text string coming from the input receiver into smaller units. They usually correspond to words or subwords, but the specific granularity may vary depending on different application scenarios (Liu et al., 2021).

Embedding is the conversion of tokens into vectors for the model after tokenization to achieve the ability to understand and generate human language. In this process, each token in the tokenization stage is mapped into a high-dimensional vector (Rothman & Gulli, 2022). The final result is a matrix, with each row representing the d-dimensional embedding of a token in the input text, and provides the basis for subsequent positional encoding (Zheng et al., 2021; Naeve et al., 2024). However, due to the architectural limitations of the model itself, LLM cannot naturally identify sequential information in sequences (Chen et al., 2021). To solve this problem, LLM ensures that the model can grasp the relative position of markers in the sequence by adding position encoding to marker embeddings (Kazemnejad et al., 2024).

Considering that the conversion layer consists of a multi-head self-attention mechanism and a feedforward neural network, the embedded information provides the core basis for the processing of the conversion layer (Reza et al., 2022; Zheng et al., 2024). Multi-head self-attention enables the model to deeply understand the complex structure of grammatical and semantic relationships in text by measuring the relative importance of each token compared to other tokens in the context (Vaswani, 2017; Xiao et al., 2022).

When the position encoding vector is generated, the system adds it to the corresponding mark embedding vector in the embedding stage. The result of this process is a set of vector sequences, where each vector not only contains the semantic information of the tags in the embedding stage, but also incorporates the relative position information in the sequence provided by the position encoding (Naveed et al., 2023; Naeve et al., 2024). The generated vector sequence (encapsulating the semantic representation of the mark and its sequence in the sequence position) into the forward pass phase (Luo et al., 2023). Each transformer's decoder receives input from the self-attention and feed-forward neural network, executes, and so on. At this stage, tokens, embeddings, and positional encodings are passed through a stack of transformer-decoders (Raiaan et al., 2024).

Softmax function to generate a probability distribution over the vocabulary for each token in the sequence. It essentially normalizes the logits so that they sum to 1, and the size of each logit determines the probability of the corresponding label.

**2.3 Smart Glasses**

Smart glasses are considered as wearable devices that superimpose digital data onto the user's real-world view (Surti & Mhatre, 2021), and their functionality usually depends on three components: optical system, processing unit, and user interface (Czuszynski et al., 2015). In smart glasses, the processing unit is considered the smart brain of the device, responsible for coordinating and managing core functions (2017). The main function of this unit is to process input data from various sensors, execute specified applications, and present the processing results to the user in the form of display (RajKumar et al., 2019). The data transfer process uses a structured approach to transfer the decoded text from the LLM to the smart glasses. Through the optical system, the processed data is superimposed into the user's real-time field of view, thereby visually displaying the processing results to the user (Xu et al., 2024)

The purpose of data parsing is to correctly interpret the data received from the previous "data transfer" stage (Novac, 2022; Soldner et al., 2022). It breaks down a string of data into smaller tokens, and these components are easier to manage and interpret by the smart glasses' computing system. Parsing is usually guided by a set of syntactic rules, which in this case will be dictated by the text formatting and display requirements of the smart glasses user interface.

The concept of "Text Formatting" in the context of smart glasses involves the integration of parsed text data into a format that is both visually appealing and user-readable (Firstenberg & Salas, 2014). Text formatting operations are a key factor in ensuring user satisfaction and system usability that involves aspects such as font size, typeface, text alignment, spacing, color, and other visual attributes that can affect readability and the overall user experience (Guo et al., 2022). Text rendering facilitates the implementation of processed and structured data on the interface of smartglasses. Acquired through text formatting, depending on the display technology of the smartglasses (Lee & Hui, 2018; Huang, 2022). In addtion, the role of the display is to successfully communicate the output of the model to the user in a readable, clear and efficient manner.

## 3. Proposed Algorithms

In contrast, mathematical formula allows systematic variation of model parameters, which facilitates the exploration of operational boundaries and conditions under which interaction models can perform optimally (Bubeck et al., 2023).

The architecture is divided into four conceptual steps from exposure to user speech: speech capture, feature extraction, acoustic modeling, and language modeling. Examining performance requires a thorough evaluation of the different components to reflect their unique contributions to the overall functionality of the ASR system.

Word Error Rate (WER) provides an overall assessment of ASR performance by quantifying ASR systems' quantifying insertions $I$, deletions $D$, and substitutions $S$ compared to human-transcribed reference texts (Ali & Renals, 2018) . The calculation takes into account the total number of words $N$ in the reference text. This key evaluation indicator is expressed as:

$$WER = (S + D + I) / N$$

- Voice capture: While there is no clear metric to evaluate the quality of voice capture, an indirect measure, the signal-to-noise ratio (SNR), can be used. A higher SNR indicates superior capture quality.

- Feature extraction: Feature extraction in ASR usually uses Mel-frequency cepstral coefficients (MFCC). While there is no direct measure of feature extraction quality, its impact can be inferred from WER. A high WER may indicate insufficient feature extraction.

- Acoustic Modeling: Acoustic model performance can be measured using the Frame Error Rate (FER), a metric that records the proportion of misclassified frames:

$$FER = \text{Incorrectly classified frames} / \text{Total frames}$$

- Language Modeling: Perplexity is a common metric for evaluating language models. A higher language model is indicated by a lower perplexity value. For a test set of N words $W1, W2, ..., WN$, perplexity is defined as:

$$\text{Perplexity} = \left(\prod_{i=1}^{N} 1/P(W_i|W_{i-1}, ..., W_1)\right)^{(1/N)}$$

The combination of the above formulas provides a comprehensive framework for critically evaluating the accuracy of ASR systems.

The coherence of speech-to-text translation of an automatic speech recognition (ASR) system is intricately linked to its ability to maintain logical and contextual continuity in the output text (Narisetty et al., 2022).

Below is a breakdown of potential metrics for the four main components of an ASR system:

- Speech Capture: High quality voice capture ensures that the ASR system has a good starting point for translation. Any misunderstanding at this stage will affect subsequent stages. While there is no specific mathematical formula to measure coherence at this stage, a high signal-to-noise ratio (SNR) will enhance the clarity of captured speech and indirectly improve coherence.
- Feature extraction: In ASR, feature extraction techniques such as Mel-frequency cepstral coefficients (MFCC) are used. Sufficient feature extraction can preserve the essential characteristics of speech, which is necessary for coherence.
- Acoustic Modeling: An acoustic model converts an acoustic signal into a sequence of phonemes or words. The quality of the model will greatly affect the coherence of the output. A low frame error rate (FER) can indicate that the acoustic model is performing well, producing coherent output.

$$FER = \textit{Misclassified frames / Total frames}$$

- Language Modeling: Coherence is primarily measured at this stage. Perplexity, the inverse probability of the test set, normalized by the number of words, can measure coherence. A lower perplexity score indicates better coherence because the language model can more accurately predict subsequent words in the sequence.

$$\textit{Perplexity} = (\textit{product from i=1 to N } (1/P(W_i|W_{i-1}, ..., W_1)))^{(1/N)}$$

Given that the nature of coherence is a more qualitative than quantitative feature, the above formulations and assessments provide indirect insight into the coherence of ASR systems.

Delay rate is critical for the seamless functioning of graphs through timely speech-to-text conversion. It is quantified as the time difference between speech input and its text representation output (Wang et al., 2022). Each stage contributes to the total latency and can be rigorously evaluated using specific metrics:

Speech capture: The delay rate of speech capture is highly hardware dependent and can be affected by factors such as microphone quality, signal-to-noise ratio, and network delay rate.

Feature extraction: Feature extraction is a computationally intensive task, so delay rate can be significant, depending on the hardware used. Fetch time "$T_{fe}$" is often used as a measure of this delay. The exact time depends on factors such as the complexity of the extraction algorithm and the computing power of the hardware.

Acoustic Modeling: Acoustic modeling that translates speech features into a sequence of phonemes or words also contributes to overall latency. This delay "$T_{am}$" can be calculated by the time it takes to run the acoustic model on the extracted features.

Language Modeling: Language modeling adds a further delay "$T_{lm}$" as it involves predicting the probability distribution of possible words following a given history of words.

The total delay "$D_{total}$" can be approximated by summing the individual delays:

$$D_{total} = T_{fe} + T_{am} + T_{lm}$$

## 4. Experiment

Since this architecture is not a model that already exists and is practiced in the current industry and academia, it is impossible to analyze the model by obtaining data. And the model represents a complex intersection of technologies involving ASR, LLMs, and smart glasses for information display. Mathematical formulas are suitable for processing and explaining large and complex data sets in the ASR-LLMs-Smart Glasses model and obtaining objective and valid results from them. Predictability is the strength of mathematical formulations that extend beyond retrospective analysis by discerning patterns and relationships in data that include predictions of future outcomes or trends in interactive loop models.

**4.1 Experimental Setup**

The researcher design of this research is based on a complex recurrent interaction model that contains three key modular components: ASR, LLMs and Smart Glasses. Its purpose is to propose ideas and analyze the linkage system operation of the above components through mathematical formula, and to simulate the collaborative operation of the interactive cycle model to realize the interactive cycle mechanism. The entire process starts with the user's voice input, which is processed by the ASR system to generate preliminary text, followed by semantic enhancement and content optimization by LLM, and finally the results are presented to the user in a visual form through smart glasses. After the user views and understands the displayed content, the system generates an audio response as the beginning of a new interaction cycle.

**4.2 Implement**

It is obvious that speech input is first recorded and converted into digital data, which is subsequently processed to extract unique acoustic features (Prabhavalkar et al., 2017; Ong et al., 2023). When these acoustic features are passed to an acoustic model, the model generates a sequence of phonemes as output (Dupont et al., 2005). These phonemes are then further processed by a language model, ultimately producing a text string that represents the best interpretation of the user's speech content (Kłosowski, 2022).

To pass text from ASR to the LLM module, you need to follow the operation process of the generative pre-training transformer, which includes seven main steps: receiving input, tokenizing, generating embeddings, adding positional encoding, performing forward propagation, calculating logarithm, Softmax is applied, and finally the results are generated through a decoder (Borgeaud et al., 2022; Katz & Belinkov, 2023). The architecture then assigns positional encodings to preserve the order of words, and the forward pass operation propagates these embeddings through the model. After obtaining the output logits, the Softmax operation converts them into probabilities, and the decoding step finally converts these probabilities back to human-readable text (Sun et al., 2021).

Refined text data from LLMs is data-fed to smart glasses. Smart glasses perform data parsing on the output text to identify the structural components of the text (Waisberg et al., 2024). The text rendering stage converts the formatted text into a form suitable for display, which is then presented to the user in the display step.

The researcher implemented the experiment using Python 3.13 IDLE. The research builds corresponding codes for ASR, LLMs and smart glasses in this architecture respectively. The code used in this research experiment is completely open and free. It has been uploaded to the GitHub repository, and the link is:

https://github.com/brucewang123456789/GeniusTrail/blob/main/Interactive%20Cycle%20Model%20-%20ASR.py

https://github.com/brucewang123456789/GeniusTrail/blob/main/Interactive%20Cycle%20Model%20-%20LLMs.py

https://github.com/brucewang123456789/GeniusTrail/blob/main/Interactive%20Cycle%20Model%20-%20Smart%20Glasses.py

## 5. Results

The code built based on this research completely implements the connection architecture of speech recognition, automatic language generation and smart glasses display, and can run successfully in the IDLE environment of Python 3.13. The experimental results are shown in Figure 2.

Figure 2 - Experimental results run through Python 3.13 IDLE

Observing the experimental results, the ASR part successfully simulated audio capture and converted it into text fragments, showing a stable modular design. However, the LLMs part had a "list index out of range" error during the decoding stage, which indicates that sequence length or index out-of-range issues may not have been properly handled when performing decoding. This error usually results from the input sequence not corresponding to the weight matrix or from inconsistencies in the processing steps. In addition, the data transmission, parsing, formatting, rendering and display modules of the smart glasses completely ran successfully, and the output results were in line with expectations, demonstrating complete message flow processing capabilities. Discussing these results, it is found that the overall architecture has potential modularity advantages, however, the LLMs part requires further optimization of the decoding logic, such as checking sequence boundaries or ensuring that input vectors match model weights. In addition, the successful execution of the smart glasses module shows that the data processing flow is relatively stable in the back-end part.

## 6. Limitation

After the acquired human speech information is converted into text, the interaction loop model needs to make integration and synchronization of ASR with LLMs. After ASR converts the user's speech into text, the output needs to be processed by llm to achieve natural language understanding and response generation. This means that the linking process between ASR and LLM technology and systems is infinitely complex and challenging. However, in current practice, real-time synchronization and data exchange between ASR and large language model systems, error handling, and management of potential delays or waiting times may encounter technical obstacles. And the response speed of the model may be slower than the real-time changes of the text generated by LLMs, which may cause high latency to affect user experience.

## 7. Summary

This research proposes an interactive cycle model centered on the "ASR-LLMs-Smart Glasses" linkage model. It is committed to promoting human-computer interaction and enhancing human capabilities through the combination of voice and vision with the model after the user wears it. The model integrates automatic speech recognition, large language model and a novel human-computer interaction architecture for smart glasses.

In order to verify the objective feasibility of the model in the absence of data due to the absence of relevant physical products, this research introduces the concept of mathematical formula to evaluate and quantify the performance of the model, which includes coherence (C), accuracy (A), Error Rate (E), Delay Rate (D) and Efficiency (Ef) metrics.

This research establishes a solid foundation for the theoretical and practical application of the ASR-LLMs-Smart Glasses model. The incorporation of this cutting-edge technology in a seamless interaction model represents a major advance in the field of human-computer interaction.